\documentclass[a4paper,11pt]{article}
\usepackage{pos}
\usepackage{floatrow}
\usepackage{enumitem}  
\newcommand{\hc}{{\rm h.c.}}
\renewcommand{\logo}{\relax}

\title{Revisiting two dark matter candidates in $S_3$-symmetric three-Higgs-doublet models}
\ShortTitle{Revisiting two dark matter candidates in $S_3$-symmetric three-Higgs-doublet models}

\author*[a]{Anton Kun\v cinas}
\author[b]{Odd Magne Ogreid}
\author[c]{Per Osland}
\author[a]{M. Nesbitt Rebelo}

\affiliation[a]{Centro de F\'isica Te\'orica de Part\'iculas, CFTP, Departamento de F\'\i sica,\\ Instituto Superior T\'ecnico, Universidade de Lisboa,\\
Avenida Rovisco Pais nr. 1, 1049-001 Lisboa, Portugal}

\affiliation[b]{Western Norway University of Applied Sciences,\\ Postboks 7030, N-5020 Bergen, 
Norway}

\affiliation[c]{Department of Physics and Technology, University of Bergen, \\
Postboks 7803, N-5020  Bergen, Norway}

\emailAdd{Anton.Kuncinas@tecnico.ulisboa.pt}
\emailAdd{omo@hvl.no}
\emailAdd{Per.Osland@uib.no}
\emailAdd{rebelo@tecnico.ulisboa.pt}

\abstract{
Models with an extended scalar electroweak sector are well motivated. Such models could accommodate a dark matter candidate if there is an additional scalar representation with a vanishing vacuum expectation value and, in addition, there are no couplings between fermions and the dark matter candidate. The most natural way to have these conditions implemented is to consider models where an underlying symmetry is imposed. Governed by this, we consider a three-Higgs-doublet model with an $S_3$ symmetry. Within this framework there are different implementations which could possibly accommodate a dark matter candidate. The family of $S_3$-symmetric three-Higgs-doublet implementations arises due to different vacua and, as a result, different minimisation conditions. In this framework the dark matter candidate falls into the class of weakly interacting massive particles. The dark matter candidate is associated with an $\mathbb{Z}_2$ symmetry which survives spontaneous symmetry breaking and is a remnant of the $S_3$ symmetry. We explore two cases, they share many aspects of the Type-I two-Higgs-doublet model plus an inert SU(2) doublet. The main difference between these two cases is the presence of an irremovable phase, which leads to CP violation in one of the implementations. The two candidate cases differ from other previously studied models with three scalar doublets by the fact that they do not allow for heavy dark matter candidates, $\mathcal{O}(500)\text{ GeV}$. Valid dark matter regions were identified as $m_\mathrm{DM} \in [52.5,\,89]~\text{GeV}$ for a model without CP violation and $m_\mathrm{DM} \in [6.5,\,44.5]~\text{GeV}$ for a model with CP violation. In the present work we refine the parameter space by applying additional checks to our previous work coming from LHC data and from indirect detection data.
}

\FullConference{%
  8th Symposium on Prospects in the Physics of Discrete Symmetries (DISCRETE 2022)\\
  7-11 November, 2022\\
  Baden-Baden, Germany
}

\begin{document}

\maketitle

\section{Introduction}

In spite of the success of the Standard Model (SM) of Particle Physics and existing cosmological observations we have a limited knowledge of the Universe~\cite{ParticleDataGroup:2022pth}. The standard cosmological model implies that a significant part of the matter of the Universe is made up of a hypothetical Dark Matter (DM). Apart from the experimental bounds and compelling evidence coming from different cosmological scales and phenomena we still do not know what DM could be. In scenarios beyond the SM it is common to invoke models with an extended scalar sector. Such extensions are appealing due to their simplicity from the mathematical point of view and the ability to deal with several shortcomings of the SM. Therefore, models with a non-minimal Higgs sector are well motivated, despite the fact that the properties of the observed Higgs boson are in good agreement with the SM predictions~\cite{10years}. In the Inert Doublet Model (IDM)~\cite{IDM} the scalar sector of the SM is enlarged by an additional copy of the SU(2) Higgs doublet, providing a DM candidate.

\begin{figure}[htb]
\floatbox[{\capbeside\thisfloatsetup{capbesideposition={right,top},capbesidewidth=6.3cm}}]{figure}[\FBwidth]
{\hspace*{-2pt}\vspace*{-3pt}\caption{Sketch of allowed DM mass ranges up to 1~TeV in various models. Blue: IDM according to Refs.~\cite{IDM_res}, the pale region indicates a non-saturated relic density. Red: IDM2 \cite{IDM2}. Ochre: 3HDM without \cite{3HDM_Z2} and with CP violation \cite{Cordero-Cid:2016krd}. Green: $S_3$-symmetric 3HDM without CP violation (R-II-1a)~\cite{Khater:2021wcx} and with CP violation (C-III-a)~\cite{Kuncinas:2022whn}.}\label{Fig:mass-ranges}}
{\includegraphics[scale=0.4]{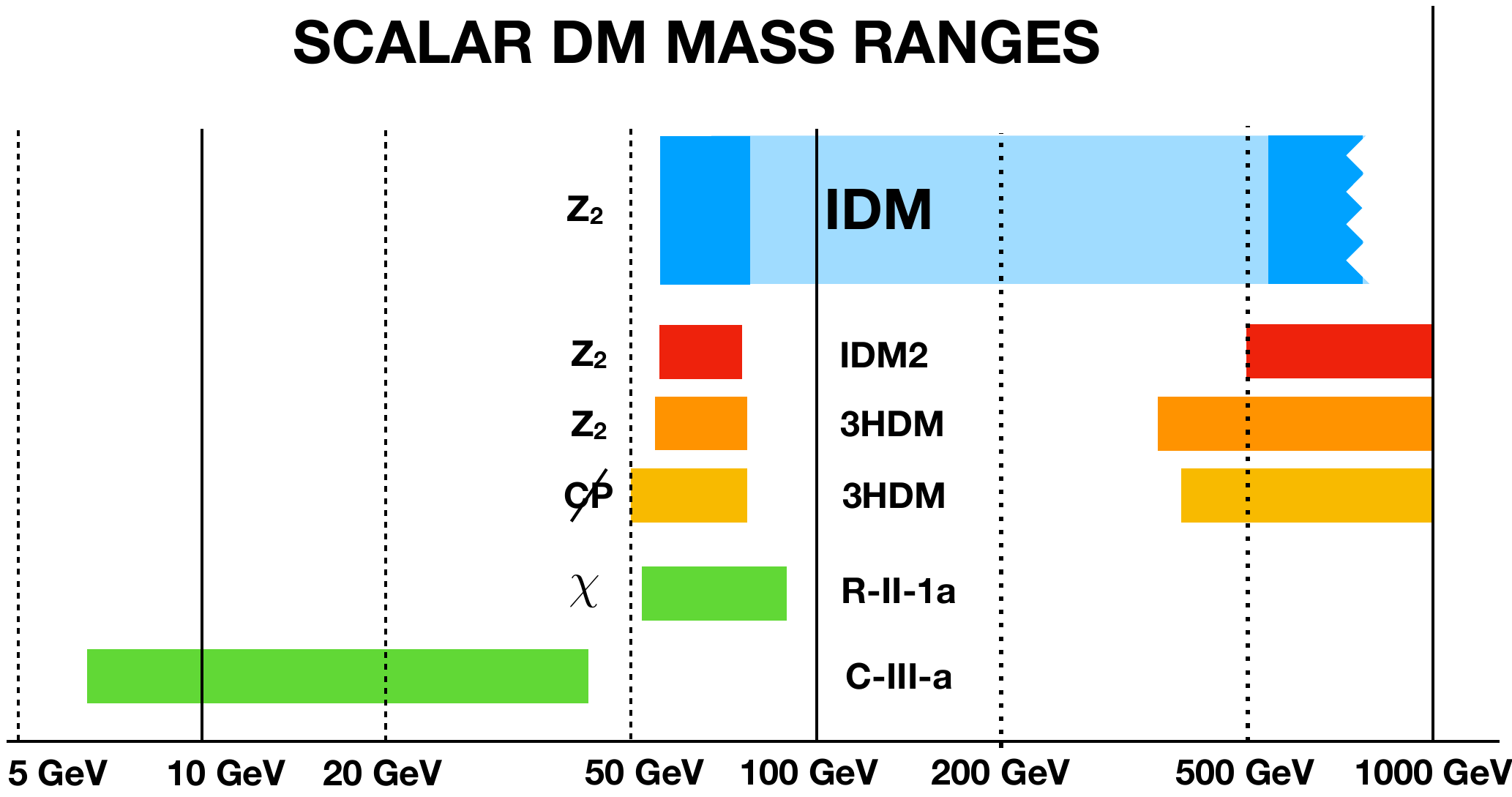}}
\end{figure}

This paper summarises two cases, R-II-1a~\cite{Khater:2021wcx} and C-III-a~\cite{Kuncinas:2022whn}, where an $S_3$-symmetric three-Higgs-doublet  model (3HDM) is assumed~\cite{Derman}. The C-III-a implementation allows for spontaneous CP violation, unlike R-II-1a. In the present review, up-to date numerical experimental bounds are used and two additional constraints coming from LHC data and from indirect DM detection data are applied. The DM candidate mass ranges of the two cases along with other models considered in the literature are presented in figure~\ref{Fig:mass-ranges}.

\vspace*{3pt}
\vspace*{-4pt}
\section{The \boldmath$S_3$-symmetric model}

The $S_3$-symmetric 3HDM potential in the doublet-singlet representation can be written as~\cite{3HDM_pot}:
\vspace*{-12pt}\begin{subequations} \label{Eq:V_S3-3HDM}
\begin{align*}
V_2=\,&{}\mu_0^2 h_S^\dagger h_S +\mu_1^2(h_1^\dagger h_1 + h_2^\dagger h_2), \\
V_4=\,&{}
\lambda_1(h_1^\dagger h_1 + h_2^\dagger h_2)^2 
+\lambda_2(h_1^\dagger h_2 - h_2^\dagger h_1)^2
+\lambda_3[(h_1^\dagger h_1 - h_2^\dagger h_2)^2+(h_1^\dagger h_2 + h_2^\dagger h_1)^2]
\nonumber \\
&+ \lambda_4[(h_S^\dagger h_1)(h_1^\dagger h_2+h_2^\dagger h_1)
+(h_S^\dagger h_2)(h_1^\dagger h_1-h_2^\dagger h_2)+\hc] 
+\lambda_5(h_S^\dagger h_S)(h_1^\dagger h_1 + h_2^\dagger h_2) \nonumber \\
&+\lambda_6[(h_S^\dagger h_1)(h_1^\dagger h_S)+(h_S^\dagger h_2)(h_2^\dagger h_S)] 
+\lambda_7[(h_S^\dagger h_1)(h_S^\dagger h_1) + (h_S^\dagger h_2)(h_S^\dagger h_2) +\hc]
\nonumber \\
&+\lambda_8(h_S^\dagger h_S)^2,
\end{align*}
\end{subequations}
here all couplings are chosen to be real, and hence CP is not violated explicitly. This  does not remove the possibility of breaking CP spontaneously; the vacuum expectation values can be complex~\cite{Emmanuel-Costa:2016vej}.

The scalar potential exhibits a $\mathbb{Z}_2$ symmetry under which $h_1 \leftrightarrow - h_1$. Therefore, in order to stabilise the DM candidate we do not need any additional symmetries. When the minimisation conditions require $\lambda_4=0$ the whole potential becomes O(2) invariant. This symmetry can be spontaneously broken by the vacuum giving rise to additional unwanted Goldstone bosons~\cite{Kuncinas:2020wrn}.

\section{Scalar dark matter candidates}

All possible implementations which could accommodate a DM candidate within the $S_3$-symmetric 3HDM were identified in Ref.~\cite{Khater:2021wcx} based on the classification of vacua of Ref.~\cite{Emmanuel-Costa:2016vej}. In total, there are three solutions with an exact $S_3$ symmetry and eight which require soft symmetry breaking of $S_3$ (in order to eliminate massless states) which could accommodate a DM candidate.

There are several possibilities to assign $S_3$ charges to fermions. The trivial approach consists of assuming that all fermions are singlets under $S_3$. In this case fermions can only couple to $h_S$, and realistic masses and mixing can be generated. The other possibility is to assume that some of the fermions are grouped into $S_3$ doublets. In this case there are seven options to assign fermions to a singlet or a pseudo-singlet representation.  Not all of the cases lead to realistic masses and mixing. 

The two cases we analysed have vacua given by $(0,w_2,w_S)$ in R-II-1a~\cite{Khater:2021wcx} or $(0,\hat w_2 e^{i \sigma},\hat w_S)$ in C-III-a~\cite{Kuncinas:2022whn}. The $\sigma$ phase is responsible for spontaneous CP violation~\cite{Emmanuel-Costa:2016vej}. One might expect to recover the R-II-1a model from C-III-a in the limit of $\sigma \to 0$. This is not the case~\cite{Kuncinas:2022whn}. The explanation is straightforward if one considers the minimisation conditions. In the C-III-a model an additional constraint arises relating two couplings, $\lambda_4$ and $\lambda_7$. As a result, these two models correspond to different regions of the parameter space, and different DM mass ranges survive. 

\section{Analysis of the models}

Both implementations are described in terms of eight input parameters. The C-III-a case is more involved than R-II-1a due to the fact that there is CP violation. In order to identify the viable DM mass region several constraints are imposed:
\begin{itemize}
\item Cut~1: perturbativity, stability, unitarity checks, LEP constraints;
\item Cut~2: SM-like gauge and Yukawa sector, $S$ and $T$ variables, $\overline B \to X(s)\gamma$ decays;
\item Cut~3: SM-like Higgs particle decays, DM relic density, direct searches;
\end{itemize}
with each subsequent constraint being superimposed over the previous one. The applied numerical bounds are taken from the PDG~\cite{ParticleDataGroup:2022pth}. We allow for a 3-$\sigma$ tolerance in relevant checks. In order to evaluate the Cut~3 constraints we used $\mathsf{micrOMEGAs}$~\cite{Belanger:2013oya}. 

The surviving parameter space of the two models projected onto the allowed mass region for the DM candidate can be seen in figure~\ref{Fig:mass-ranges}. The mass regions for DM in R-II-1a and C-III-a differ from the IDM and other 3HDMs. There is no high mass region for these two cases. In this region the relic density could be maintained by suppressing annihilation via intermediate neutral scalar bosons and into a pair of neutral scalar bosons while also requiring near mass degeneracy among the scalars of the inert sector. In neither of the cases it is possible to maintain weak portal couplings, e.g., in the SM limit the portal coupling of the R-II-1a model scales like $\sim m_\mathrm{DM}^2/v^2$. In the C-III-a case the situation is even more complicated since it is not possible to have a near mass degeneracy between the neutral inert states. A gap of around 70 GeV develops for masses above 300 GeV. 

\begin{figure}[h]
\floatbox[{\capbeside\thisfloatsetup{capbesideposition={right,top},capbesidewidth=6.1cm}}]{figure}[\FBwidth]
{\vspace*{-8pt}\caption{The spin-independent DM-nucleon cross-section compatible with XENON1T~\cite{Aprile:2018dbl} data at 90\% C.L. (yellow band). The dashed lines represent future sensitivities of XENONnT~\cite{XENON:2020kmp} (orange) and LUX-ZEPLIN~\cite{LZ:2022ufs} (green). The points (light grey: C-III-a, grey: R-II-1a) represent cases that satisfy Cut~3. The red line corresponds to an approximate neutrino floor.}\label{Fig:DM_DD}}
{\includegraphics[scale=0.28]{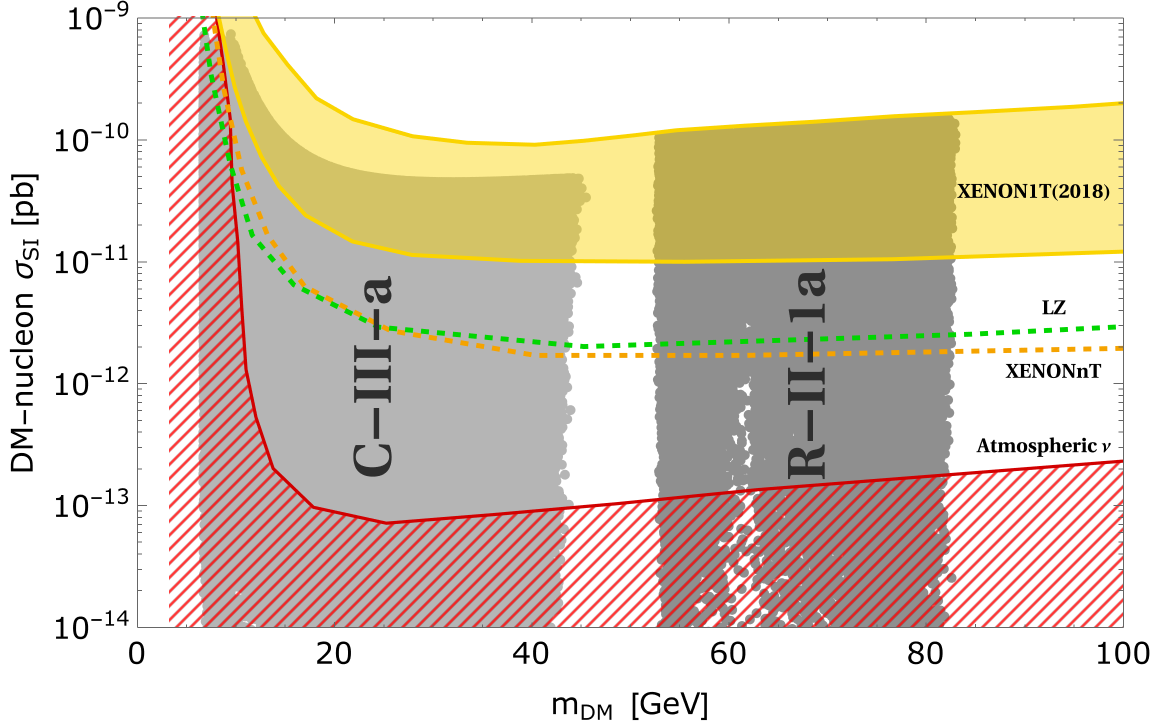}}
\end{figure}

Another interesting property of the C-III-a implementation is that the allowed DM masses are below 50 GeV. This region, in several models, is mostly not consistent with simultaneously fitting the relic density and other relevant constraints. In the C-III-a case the light DM region is compatible with the direct DM detection constraints, see figure~\ref{Fig:DM_DD}. The dominant decay channel for scalars, except for the SM-like Higgs boson, is into states with at least one DM candidate. Such processes would be accompanied by large missing transverse momentum in the detector. The SM-like Higgs portal coupling can be very weak, which allows for the SM-like Higgs branching ratio to the DM scalar states to be acceptable within the experimental bounds for DM states as light as $\mathcal{O}(\text{GeV})$.

\vspace*{6pt}
\vspace*{-3pt}
\section{Additional constraints applied to our previous work}

We applied additional constraints beyond those presented in Refs.~\cite{Khater:2021wcx,Kuncinas:2022whn}, which could be considered as Cut~4:
\begin{itemize}
\item LHC searches implemented in $\mathsf{HiggsTools}$~\cite{Bahl:2022igd}; 
\item Indirect DM detection constraints;
\item Direct DM detection constraints from the currently running experiments;
\item Significantly improved results on the SM-like Higgs boson decays into invisible states.
\end{itemize} 

\begin{figure}[h]
\begin{center}
\includegraphics[scale=0.23]{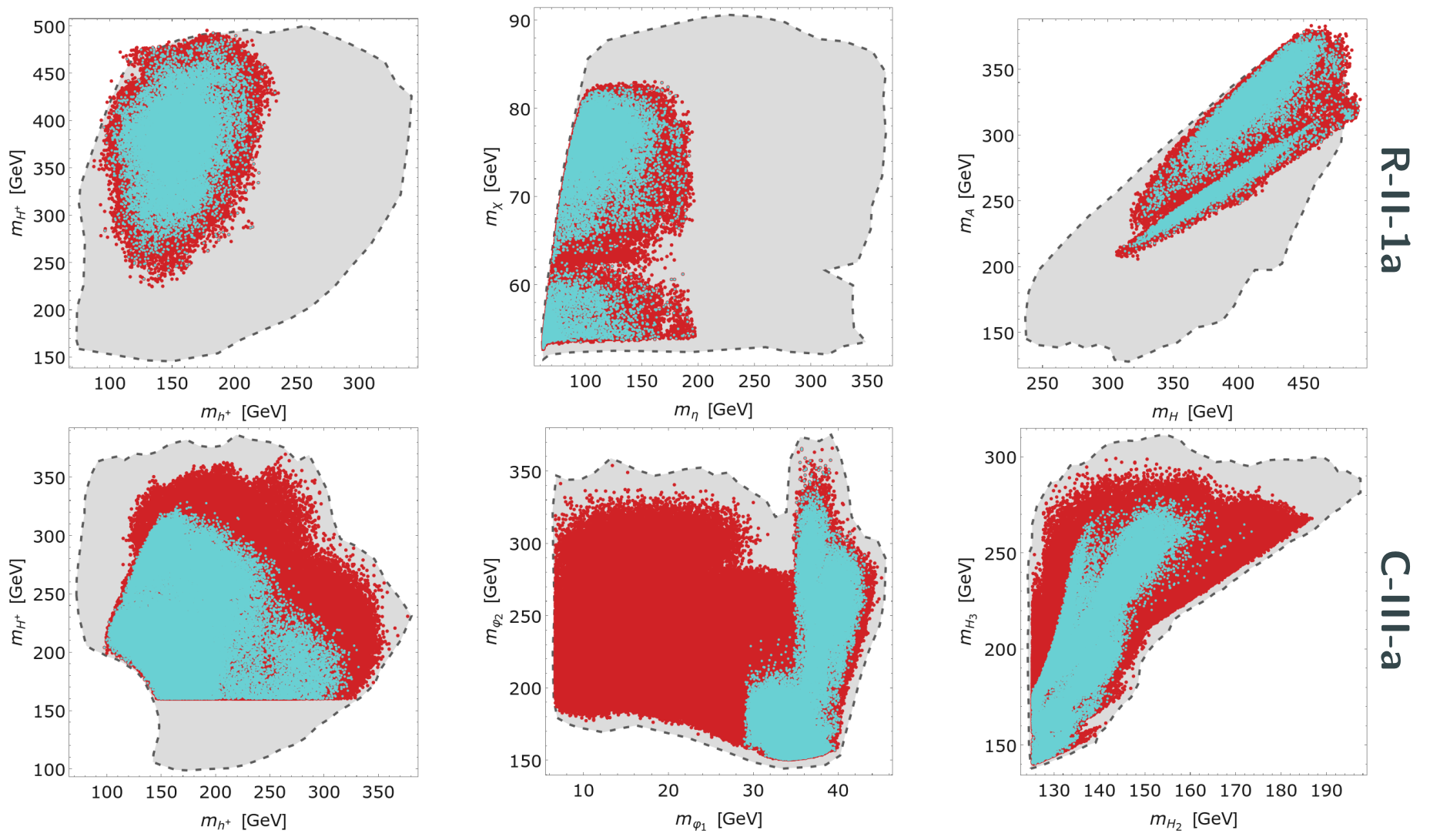}
\end{center}
\vspace*{-8mm}
\caption{Scatter plots of masses that satisfy different sets of successive Cuts. Left column: the charged sector. Middle column: the inert neutral sector. Right column: the active heavy neutral sector. The grey region satisfies Cut~3. The red region relies on the $\mathsf{HiggsTools}$ framework. The cyan region accommodates several constraints: indirect DM detection, currently running relevant direct DM detection experiments and assumes the branching ratio of the SM-like Higgs portal to DM to be within $\mathrm{Br}(h \to \text{inv.}) \leq 0.1$.}
\label{Fig:Cut4}
\end{figure}

First of all, we implemented additional checks using the $\mathsf{HiggsTools}$ framework, which utilises both $\mathsf{HiggsBounds}$~\cite{Bechtle:2020uwn} and $\mathsf{HiggsSignals}$~\cite{Bechtle:2020pkv} public codes. No significant exclusion of the parameter space imposed by $\mathsf{HiggsSignals}$ was found. However, the available parameter space in R-II-1a and C-III-a was significantly reduced by the $\mathsf{HiggsBounds}$ check. One of the more important experimental constraints not taken into consideration in the initial analysis were decays $t \to H^\pm b$. The surviving parameter space of the scalar masses is presented in figure~\ref{Fig:Cut4}.

\begin{figure}[h]
\floatbox[{\capbeside\thisfloatsetup{capbesideposition={right,top},capbesidewidth=6.05cm}}]{figure}[\FBwidth]
{\vspace*{12pt}\caption{DM self-annihilation cross-section as a function of the DM mass. The yellow band represents bounds at 90\% C.L. compatible with observation of 20 dwarf spheroidal galaxies (dSphs). The red line represents Fermi-LAT assuming the NFW profile with $\rho = 0.3\text{ GeV/cm$^3$}$. The dashed lines represent expectations from future sensitivities of Fermi-LAT (green) and probes by the Cherenkov Telescope Array (CTA) (purple). The light grey points represent C-III-a and the grey points R-II-1a cases that survive $\mathsf{HiggsTools}$ checks.
}\label{Fig:InDD}}
{\includegraphics[scale=0.28]{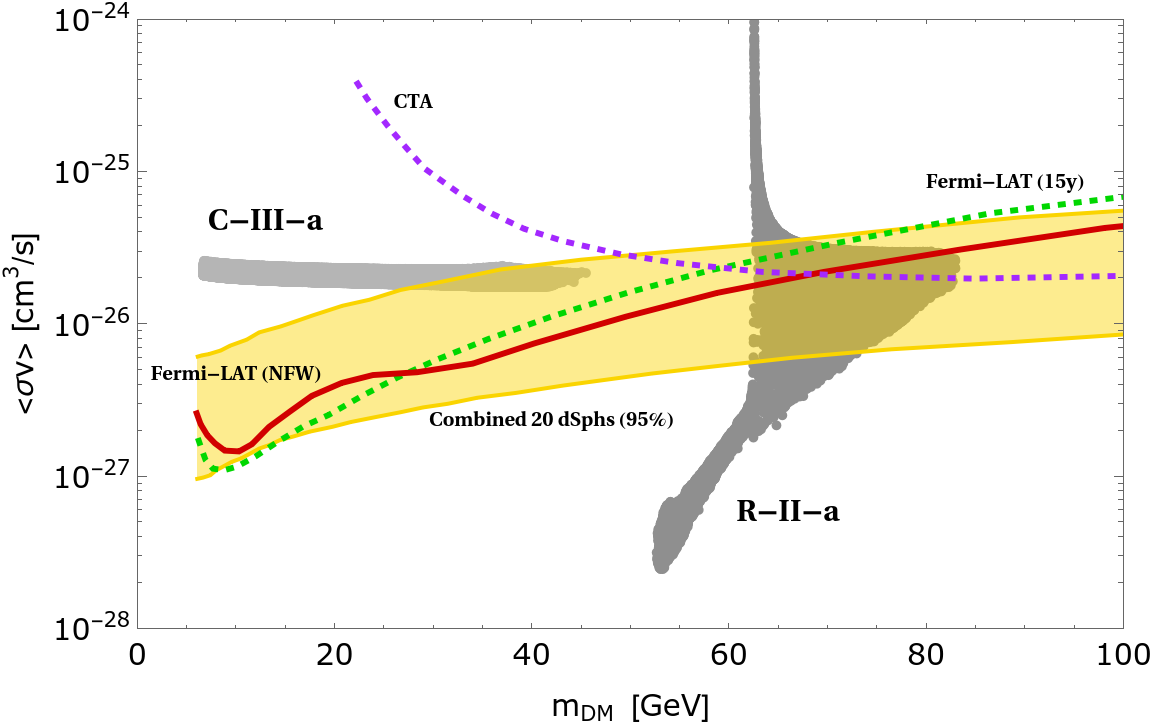}}
\end{figure}

The light scalar DM mass region is strongly constrained~\cite{Hess:2021cdp}, ruling out the canonical cross-section $10^{-26} \text{cm$^3$/s}$~\cite{Fermi-LAT:2015att}. Applying the experimental DM annihilation bounds to the two cases is not simple. The cases are analysed at tree level.
In the C-III-a case the dominant annihilation channel is into a $b \bar b$ pair with $\text{Br} \geq 0.8$. The C-III-a model would be ruled out if the NFW~\cite{Navarro:1995iw} DM distribution profile ($\rho = 0.3\text{ GeV/cm$^3$}$) would be taken, however such constraints are not fully established. In the R-II-1a case there are two dominant annihilation channels, either into a $b \bar b$ pair or a $W^+ W^-$ pair. In both cases the dominant branching ratio can be as low as 0.38.
Due to these limitations we allow for the DM self annihilation cross-section to be within a generous one order of magnitude of the experimental bounds. Results are presented in figure~\ref{Fig:InDD}.

\begin{figure}[h]
\floatbox[{\capbeside\thisfloatsetup{capbesideposition={right,top},capbesidewidth=7.2cm}}]{figure}[\FBwidth]
{\vspace*{-25pt}\caption{Probing the CP properties of the SM-like Higgs boson-fermion coupling for the C-III-a model. The neutral scalar-fermion interaction terms $(\phi^0 \overline{\psi_f} \psi_f)$ can be read from the Yukawa Lagrangian,\\
$
\mathcal{L} \supset-\frac{m_f}{v} \overline{\psi_f} \phi^0 \left(\kappa_f+i \gamma_5 \tilde{\kappa}_f\right) \psi_f.
$
The CP-odd/even components are within 2-$\sigma$ of the combined LHC data. The current best fit of the LHC data is depicted in the lower-left panel.
}\label{Fig:CP_Hff}}
{\includegraphics[scale=0.4]{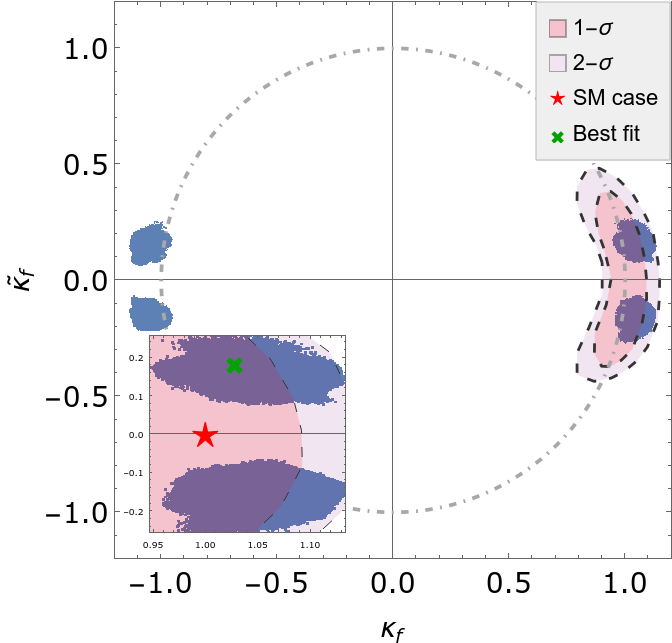}}
\end{figure}

In the C-III-a case the SM-like Higgs boson couples to fermions as a CP-indefinite state. It is possible to compare the CP-indefinite couplings against the current LHC bounds~\cite{LHC_CP}. The CP-odd/even couplings for C-III-a are presented in figure~\ref{Fig:CP_Hff}. The current LHC data suggests the positive CP-even sign of the Yukawa couplings ($\kappa_f$). ATLAS and CMS measure the Higgs-fermion couplings using different approaches. The CP-odd Yukawa couplings ($\tilde{\kappa}_f$) of the surviving parameter space after applying Cut~4 is well within the current measurements (2-$\sigma$).  

\vspace*{25pt}

After applying Cut~4 the allowed DM mass region of the R-II-1a model remained almost unchanged, from $m_\mathrm{DM} \in [52.5,\,89]~\text{GeV}$ it was reduced to $m_\mathrm{DM} \in [53,\,83]~\text{GeV}$. However, the overall parameter space was drastically constrained by $\mathsf{HiggsBounds}$. The case of the C-III-a implementation is more interesting since it allowed for rather light DM candidates. Before applying Cut~4 the allowed DM mass region was shown to be $m_\mathrm{DM} \in [6.5,\,44.5]~\text{GeV}$. Such light states could be further probed by both collider and indirect DM detection searches. The collider searches turned out to be less significant in limiting the available parameter space due to the phenomenology of the model (decay channels). Yet, the experimental bounds from the indirect DM detection could completely rule out the C-III-a case, assuming some specific DM halo distribution profiles. If generous bounds are applied the surviving DM mass region is reduced to $m_\mathrm{DM} \in [29,\,44]~\text{GeV}$.

\acknowledgments
PO~is supported in part by the Research Council of Norway. The work of AK and MNR was partially supported by Funda\c c\~ao para a Ci\^encia e a Tecnologia (FCT, Portugal) through the projects CFTP-FCT Unit UIDB/00777/2020 and UIDP/00777/2020, CERN/FIS-PAR/0002/2021 and CERN/FIS-PAR/0008/2019, which are partially funded  through POCTI (FEDER), COMPETE, QREN and EU. Furthermore, the work of AK has been supported by the FCT PhD fellowship with reference UI/BD/150735/2020. We also thank the University of Bergen and CFTP/IST/University of Lisbon, where collaboration visits took place. 

\bibliographystyle{JHEP}

\bibliography{ref}

\end{document}